\documentclass[aps,pra,twocolumn,showpacs,amsmath,amssymb,superscriptaddress,longbibliography, superscriptaddress]{revtex4-1}

\usepackage{graphicx}
\usepackage{verbatim}
\usepackage{epstopdf}
\usepackage{bm}
\usepackage{times}
\usepackage{comment}
\usepackage{dsfont}
\usepackage{color}
\usepackage[english]{babel}
\definecolor{blue}{RGB}{0,0,255}
\usepackage[colorlinks,citecolor=blue,linkcolor=blue,urlcolor=blue]{hyperref}
\pdfoutput=1

\def\BE{\begin{equation}}
\def\EE{\end{equation}}
\def\BY{\begin{eqnarray}}
\def\EY{\end{eqnarray}}
\def\BI{\begin{itemize}}
\def\EI{\end{itemize}}

\def\({\left (}
\def\){\right)}
\def\[{\left [}
\def\]{\right]}
\def\<{\langle}
\def\>{\rangle}

\def\BA{\begin{array}}
\def\EA{\end{array}}

\def\dd{\delta}

\def\t{\tau}

\def\+{\dag}
\def\8{\infty}

\def\={\approx}

\def\->{\rightarrow}

\newcommand{\ud}{\,\mathrm{d}} 

\def\um{$\rm{\mu m}$}
\def\um1{$\rm{\mu m}^{-1}$}

\usepackage{svg}

\newcommand{\W}{\Omega}
\newcommand{\w}{\omega}
\newcommand{\CR}{\sqrt{\mathcal{R}}}
\newcommand{\CT}{\sqrt{\mathcal{T}}}
\newcommand{\CE}{\mathcal{E}}

\usepackage{cancel}
\usepackage{enumitem}
\graphicspath{{images/}}

\allowdisplaybreaks[1] 

\usepackage{indentfirst}
\usepackage[normalem]{ulem}  

\usepackage[breakable]{tcolorbox}

\begin{document}

\title{
Limits of Perturbation Theory for Multimode Light Propagation in Dispersive Optical Cavities
}

\author{Kirill Tikhonov}
\email{tikhonov.kyril@gmail.com}
\affiliation{St. Petersburg State University, 7/9 Universitetskaya Nab., 199034 St. Petersburg, Russia}

\author{Danil Malyshev}
\affiliation{St. Petersburg State University, 7/9 Universitetskaya Nab., 199034 St. Petersburg, Russia}

\author{Valentin Averchenko}
\email{valentin.averchenko@gmail.com}
\affiliation{St. Petersburg State University, 7/9 Universitetskaya Nab., 199034 St. Petersburg, Russia}

\date{\today}

\begin{abstract}
Temporal modes of quantum light pulses is a promising resource for modern quantum technologies, driving advancements in quantum computing, communication, and metrology. Precise control and manipulation of these modes remain critical challenges, particularly in systems where nonlinear multimode dynamics interact with dispersion effects. In this work, we focus on the role of group velocity dispersion (GVD) within optical cavities—a phenomenon traditionally viewed as detrimental but increasingly recognized as a versatile tool for quantum light manipulation.
We present a perturbation-theory-based approach to analyze GVD effects in a synchronously pumped dispersive cavity. By comparing perturbative solutions to rigorous steady-state results, we establish the validity region of the perturbative approach and assess its limitations in multimode systems. Our study identifies key parameters governing the breakdown of perturbation theory, such as mode order, dispersion strength, and cavity decay rates.
\end{abstract}

\maketitle

\section{Introduction}
Temporal modes of quantum light pulses have emerged as a fundamental resource in modern quantum technologies \cite{fabre_modes_2020}, enabling breakthroughs in linear-optical quantum computing, boson sampling, quantum communication, and quantum metrology. However, harnessing their full potential poses several critical challenges. Among these challenges,  precise control and manipulation of the modes are imperative. This requirement remains at the forefront of current research efforts in quantum optics and quantum photonics. 

Group velocity dispersion (GVD) is a fundamental phenomenon that arises in both nonlinear dielectric media—where multimode quantum light is generated—and within discrete components of optical circuits. It plays a critical role in shaping the temporal and spectral properties of propagating light, influencing applications ranging from ultrafast optics to quantum communications. Conventionally, GVD has been regarded as a detrimental effect, as it distorts the temporal profiles of optical modes and induces undesirable crosstalk. However, recent advancements have demonstrated the potential of GVD as a versatile and powerful tool for precise quantum light manipulation. Specifically, GVD can be utilized to separate overlapping temporal modes by transforming them into distinct, non-overlapping time bins. This is achieved through a sequential temporal phase modulation and dispersion - a technique that has been successfully demonstrated for temporal Hermite-Gaussian (HG) modes \cite{brecht_demonstration_2014,ashby_temporal_2020}. Furthermore, GVD allows one to tailor the frequency spectrum of high-gain parametric down-conversion using an SU(1,1) interferometer \cite{lemieux_engineering_2016}. The process involves the selection  of a narrow band and its amplification by a high-gain parametric amplifier. Additionally, time-lenses based on GVD and incorporated into Mach-Zehnder interferometers facilitate temporal mode selection through the accumulation of the temporal Gouy phase \cite{horoshko_interferometric_2024}. Finally, the deliberate employment of GVD engineering opens new perspectives for enhancing resoultion of intracavity  spectroscopy \cite{cygan_cavity_2021}. 

The aforementioned examples demonstrate that the study of dispersion in nonlinear optical circuits constitutes is a pivotal research objective. This task becomes particularly challenging in systems where nonlinear multimode dynamics arise within a cavity, as accounting for dispersion effects introduces significant complexity. Among existing theoretical frameworks, the analytical Bloch-Messiah decomposition stands out as one of the most advanced methods for treating such a systems \cite{gouzien_morphing_2020}. 
In our previous work \cite{averchenko_effect_2024}, we addressed this challenge by introducing a perturbation-theory-based approach and applied it up to second order to study the dynamics of a synchronously pumped optical parametric oscillator (SPOPO). The core idea relies on treating weak dispersion as a small perturbation to the system. In the present work, we analyze this methodology in details by establishing the validity region of the perturbative approach using the test problem of a synchronously pumped dispersive cavity. 

\section{Propagation of a Classical Optical Pulse in the Dispersive Cavity}
\begin{figure}[htbp]
    \centering
    \includegraphics[width=0.60\columnwidth]{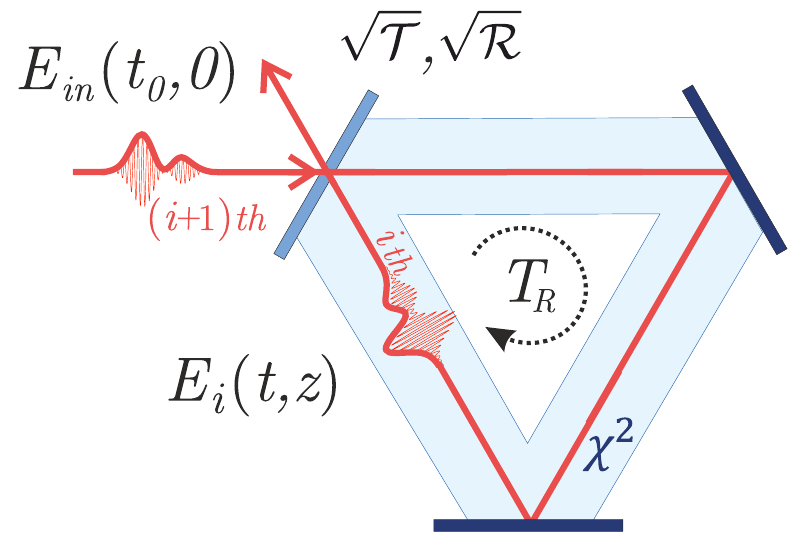}
    \caption{A ring cavity of length $L$ with a dispersing medium and a light pulse $E_i(t, z)$ propagating in it. The cavity is pumped with a pulsed light field  $E_{in}(t,0)$ through the input-output mirror with reflection coefficient $\CR$ and transmission coefficient $\CT$. The pulse undergoes a round-trip within the cavity, with a time $T_R$.  
    } 
    \label{Fig: cavity}
    \end{figure}
Our analysis begins with the canonical problem of optical pulse propagation within a dispersive ring cavity. The resonator is filled with a nonlinear $\chi^2$-medium. The system is coherently pumped by a periodic train of identical pulses, each with a carrier frequency $\w_0$. The period of the incoming pulse train is precisely matched to the cavity's round-trip time $T_R$. This resonant condition ensures that each new input pulse coherently interferes with the intracavity pulse circulating within the resonator and amplifies it.
Since the temporal duration $\tau_s$ of each pulse is substantially shorter than both the inter-pulse separation and the round-trip time $T_R$ ($\tau_s \ll T_R$)  individual pulses do not overlap temporally inside the resonator. Consequently, we can model the system dynamics by tracking the evolution of a single, representative pulse over multiple round trips within the cavity. The relationship between the electric field amplitude $E_i(t, z)$ of the pulse after its $i$th round-trip and its corresponding spectral amplitude $\CE_i(\omega_s, z)$ is governed by the Fourier transform:
\begin{align}
E_i(t, z) &= \int_{0}^{\infty} d \w_s \, \CE_i(\w_s, z) e^{-i\w_s t} + c.c., 
\label{eq: sec 1: temporal amplitude}
\end{align}

After each round-trip, the spectral amplitude of the pulse at the input mirror $(z=0)$ is changed. Specifically, $\CE_i(\omega_s, 0)$ is determined by taking the spectral amplitude from the previous round-trip, $\CE_{(i-1)}(\omega_s, 0)$, propagating it through the cavity once, and adding the contribution of the incoming pump field $\CE_{in}(\omega_s)$
\begin{align}
\CE_i(\w_s, 0) &=\CR\, \CE_{(i-1)}(\w_s, 0) e^{i k(\w_s) L} \nonumber\\
&+ \CT\, \CE_{in}(\w_s). 
\label{eq: sec 1: spectral amplitude}
\end{align}
where $\CR$ and $\CT$ are the reflection and transmission coefficients of the iput-ouput mirror and $L$ is the length of the cavity.  $k(\w_s)$  represents the dispersion relation for the optical system, defining the dependence of the wavenumber $k$ on the angular frequency $\w_s$.

In the following, we determine the intracavity stationary field amplitude using multiple approaches. One method provides an exact solution, while another method employs perturbation theory to approximate the solution. By comparing the two, we delineate the validity range of the perturbative solution.

\section{Rigorous steady-state solution}\label{Sec: rigorous solution}
Under the steady-state conditions, the intracavity light field attains a time-invariant equilibrium. Consequently, the dependence on the pulse number in Eq. \eqref{eq: sec 1: spectral amplitude} can be omitted. The spectral amplitude of the circulating pulse is then given by the following expression:
\begin{align}
\CE_{\text{cav}}(\w_s) &=   \frac{\CT \CE_{in}(\w_s)}{ 1- \CR\, e^{i k(\w_s) L}}.
\label{eq: sec 1: steady-state initial form}
\end{align}
The dispersion relation $k(\w_s)$ is expanded as the Taylor series around the carrier frequency  $\w_0$:
\begin{align}
k(\w_0 + \w) &= k(\w_0) + k'(\w_0) \w \nonumber\\
&+ \frac{k''(\w_0)}{2!} \w^2 + \frac{k'''(\w_0)}{3!} \w^3 + \ldots 
\label{eq: sec 1: k expansion}
\end{align}
where $\w$ is the relative frequency and $\w_s=\w_0+\w$. 
To simplify the analysis, we assume that the pulse's carrier frequency is resonant with the cavity mode, i.e.  $k(\w_0)L=2\pi n$ for $n\in\mathbb{N}$. This resonance condition nullifies the first term in the propagation phase expansion after substituting sum \eqref{eq: sec 1: k expansion} into equation \eqref{eq: sec 1: steady-state initial form}. Furthermore, the second term, representing a constant group delay, can be omitted as it does not alter the pulse shape. To exclusively model the influence of GVD, we retain the third term, proportional to $k''(\w_0)$, and truncate all subsequent higher-order dispersion terms, assuming:
\begin{align}
\Bigg|\frac{k'''(\w_0)}{3!} \w^3\Bigg| \ll \Bigg|\frac{k''(\w_0)}{2!} \w^2\Bigg|.
\label{eq: sec 1: GVD gg TOD}
\end{align}
This inequality is valid if the spectral power density of the pulse is concentrated in the specified frequency range, i.e. $|\w| \ll 3k''(\w_0)/k'''(\w_0)$.
Assumption \eqref{eq: sec 1: GVD gg TOD} possesses broad applicability, remaining valid across a range of experiments, e.g. for \cite{Roslund2014}.  In this work a multimode squeezed light generation utilized a Bismuth Borate (BiB$_3$O$_6$) crystal of length $L=2$ mm inside an optical cavity. At the carrier wavelength of $\lambda_0=795$ nm ($\w_0/2\pi=374.75$ THz), the crystal exhibited a GVD of $k''(\omega_0)=136$ fs$^2$/mm. Our analysis, based on the Sellmeier equation (Appendix \ref{App: Sellemeir equation}), further yields a Third-Order Dispersion (TOD) coefficient of $k'''(\omega_0)=1644$ fs$^3$/m.
Substituting these values into condition \eqref{eq: sec 1: GVD gg TOD} reveals that it requires the relative frequency to satisfy $|\w/2\pi| \ll 250$ THz. In terms of absolute signal frequency, this corresponds to a broad range of $( 150\ll \w_s/2\pi \ll 500) $ THz, or equivalently, $(600 \ll \lambda_s \ll 2000)$ nm. The supermode spectrum generated in \cite{Roslund2014}, confined to the narrow band of $785-805$ nm, lies well within this validated range. Therefore, the condition \eqref{eq: sec 1: GVD gg TOD} is fully satisfied, confirming the assumption's robustness in this experimental context.
Further, by retaining only the GVD term in the exponent's argument in \eqref{eq: sec 1: steady-state initial form} and invoking the small dispersion approximation, all higher-order terms can be omitted, leading to the following series expansion: 
\begin{align}
\text{exp}\Big({i \frac{k''(\w_0)}{2} \w^2 L}\Big) \approx 1 + i\frac{k''(\w_0)}{2} \w^2 L. \label{eq: sec 1: exp expansion}
\end{align}
Substituting \eqref{eq: sec 1: exp expansion} into \eqref{eq: sec 1: steady-state initial form} yields
\begin{align}
\CE_{\text{cav}}(\w) =   \frac{1}{ 1- \CR \(1 + i \frac{k''(\w_0)}{2} \w^2 L \) } \CT \CE_{in}(\w).
\end{align}
To identify the leading-order terms and simplify the analysis, the derived expression is expanded into its Maclaurin series::  
\begin{align}
    \CE_{\text{cav}}(\tilde \w) &=  \sum_{k=0}^{\infty}\Bigg(i\frac{N_\gamma}{N_D} \tilde \w^2\Bigg)^k \frac{\CT}{1-\CR} \CE_{\text{in}}(\tilde\w), 
    \label{eq: sec: 1 the total field}
\end{align}
where $\tilde\w = \w\tau_s$ is the dimensionless relative frequency, $\t_s$ is the characteristic duration of input pulses, $\gamma$ is the decay rate, $N_\gamma = (\gamma T_R)^{-1} = \sqrt{\mathcal{R}}/(2(1-\sqrt{\mathcal{R}}))$ is the number of round trips  after which the intensity of the light field decreases $e$-times, $N_D = \tau_s^2/(k''(\w_0)L)$ defines the number of round-trips after which a pulse broadens by a factor of $\sqrt{2}$. It represents the ratio of the dispersive length $L_D$ to the cavity length $L$, that is, $N_D=L_D/L$. Thus, inequality $N_\gamma/N_D<1$ signifies that a pulse  exits the cavity before dispersion can substantially alter its temporal profile.
Equation \eqref{eq: sec: 1 the total field} provides the steady-state solution for the light pulse amplitude, derived under the assumption of the small GVD within the cavity. For the subsequent consideration, we now express this solution as a series expansion in the basis of the Hermite-Gaussian modes $\{s_n(\tilde\omega)\}$:
\begin{align}
	s_n(\tilde\w) = i^n \Big( \sqrt{\pi} 2^n n!  \Big)^{-\frac{1}{2}} H_n(\tilde\w) e^{- \tilde\w^2/2 }, \label{eq: sec: 1 Hermite-Gaussian modes}
\end{align}
where $H_n(\tilde\W)$  is the $n$th order Hermite polynomial. 
In this basis, the amplitude of the $m$th mode entering the cavity is given by the overlapping integral:
\begin{align}
    \alpha_m^{\text{in}} = \frac{\CT}{1-\CR} \int \ud\tilde\w \; \CE_{\text{in}}(\tilde \w)s_m(\tilde \w).
\end{align}
At the same time the amplitude of the $n$th  circulating  mode is determined by:
\begin{align}
    \alpha_n^{\text{cav}} = \int \ud\tilde\w \; \CE_{\text{cav}}(\tilde \w)s_n(\tilde \w).
\end{align}
Consider that the input pulse is a Hermite-Gaussian mode with spectral profile  $s_p(\tilde\w)$. This profile is substituted into the field expansion \eqref{eq: sec: 1 the total field}. The coupling to the intracavity mode with the spectral profile $s_q(\tilde\w)$ is then determined by calculating the overlapping integral with each term of the series. In particular, the overlapping integral with the $3$rd term ($k=2$) in expansion \eqref{eq: sec: 1 the total field} is proportional to
\begin{align}
    &\int\ud\tilde \w s^*_q \tilde \w^4(\tilde \w) \CE_{in}(\tilde \w) \propto \int \ud\tilde\w s^*_q(\tilde \w) \tilde \w^4 s_p(\tilde \w) \nonumber\\
    &=  \int \ud\tilde \w s^*_q(\tilde \w) \tilde \w^2 \int\ud\tilde \w' \dd(\tilde \w-\tilde \w') \tilde \w'^2 s_p(\tilde \w')\nonumber\\
    & = \sum_m \int \ud \tilde \w s^*_q(\tilde \w) \tilde \w^2 s_p(\tilde \w) \int\ud\tilde \w' s_q^*(\tilde \w') \tilde \w'^2 s_p(\tilde \w') \nonumber\\
    &= \sum_m O_{qm} O_{mp},
\label{eq: sec 1: the coupling coefficients}
\end{align}
where we used the completness of $\{s_n(\tilde\w)\}$, that is $\dd(\tilde \w-\tilde \w') \\= \sum_m s_m(\tilde \w) s_m^*(\tilde \w')$. This derivation can be extended to the higher-order terms in  expansion \eqref{eq: sec: 1 the total field} presented above.
The dimensionless coupling coefficient $O_{nm}$ between the $n$th and $m$th Hermite-Gaussian modes is calculated as follows: 
\begin{align}
     O_{nm}&= \int \ud \tilde \w s^*_n(\tilde \w) \tilde \w^2 s_m(\tilde \w) 
     =-\Big(n+\frac{1}{2}\Big) \delta_{n m}\nonumber\\
 &- \frac{\sqrt{(n-1)n}}{2} \delta_{n m+2}  
 -  \frac{\sqrt{(n+1)(n+2)}}{2} \delta_{n m-2}.
\label{eq: sec 1: coefficients Onm}
\end{align}

Therefore, when the input pulse is a $k$-th order Hermite-Gaussian mode with a spectral profile $s_k(\tilde\omega)$, its amplitude $\alpha_k$ inside the cavity is modified by GVD according to:
\begin{align}
    \alpha_k^{\text{cav}} &= \Bigg(1 -i \frac{N_\gamma}{N_D} O_{kk}- \Bigg(\frac{N_\gamma}{N_D} \Bigg)^2 \sum_{m} O_{km}  O_{mk}  \nonumber\\    &+ i \Bigg(\frac{N_\gamma}{N_D} \Bigg)^3 \sum_{m,n} O_{km} O_{mn}   O_{nk}+...\Bigg) \alpha_k^{in}.
\label{eq: sec 1: amp_k cav - amp_k in}    
\end{align}
At the same time, the amplitude $\alpha_l^{\text{cav}}$ of any other mode (where $l\neq k$) is
\begin{align}
    \alpha_l^{\text{cav}} &= \Bigg( - i\frac{N_\gamma}{N_D}   O_{lk}  -\Bigg(\frac{N_\gamma}{N_D} \Bigg)^2 \sum_{m}  O_{lm} O_{mk} \nonumber\\
    &+ i  \Bigg(\frac{N_\gamma}{N_D} \Bigg)^3 \sum_{n,m}  O_{ln} O_{nm} O_{mk}+...\Bigg)\alpha_k^{in}.
\label{eq: sec 1: amp_l cav - amp_k in}    
\end{align}
Expressions \eqref{eq: sec 1: amp_k cav - amp_k in} and \eqref{eq: sec 1: amp_l cav - amp_k in} describe the transfer of photons from the $k$th mode to other modes via dispersive coupling inside the cavity. These expressions can be generalized for a multimode input as follows:
\begin{align}
    \boldsymbol{\alpha}^{\text{cav}} = \sum_{M=0}^{\infty}\Bigg( - i\frac{\hat N_\gamma} {N_D} \hat O\Bigg)^M \boldsymbol{\alpha}^{\text{in}}
\label{eq: sec 1: the matrix solution}
\end{align}
Here, $\boldsymbol{\alpha}^{\text{in}}$ is the vector of input mode amplitudes entering the optical cavity, and $\boldsymbol{\alpha}^{\text{cav}}$ is the vector of the mode amplitudes circulating inside it. The matrix $\hat O$ has elements $O_{nm}$, and $\hat N_\gamma$ is a diagonal matrix where each element $N_{\gamma_n}$ represents the number of round trips it takes for the intensity of a specific optical mode to decrease $e$-times  of its initial value.

The series solution presented in \eqref{eq: sec 1: the matrix solution} is made possible through the application of the Maclaurin series expansion \eqref{eq: sec: 1 the total field}. This expansion requires that any $n$th mode meet the following condition:
\begin{align}
    \Bigg|\frac{N_{\gamma_n}}{N_D}O_{nn}\Bigg| <1. 
\label{eq:  sec 1: main condition}    
\end{align}
This inequality is based on the property that for all $n>1$, the diagonal entries of the matrix $\hat{O}$ dominate the non-zero off-diagonal ones, satisfying $O_{nn} > O_{n\,(n+2)} = O_{(n+2)\,n}$. For the specific case of $n=1$, the diagonal element $O_{11}$ must be replaced with $O_{13}$ (or its symmetric counterpart $O_{31}$) to formulate the inequality correctly. Physically, the diagonal element $O_{nn}$  represents the dispersion of the $n$th mode, which defines its fundamental spectral characteristics.

\section{Perturbation theory solution of Heisenberg-Langevin equation}\label{Sec: semiclassical approach}

We now turn to a pulsed description of field propagation in a dispersive cavity. The physical configuration remains as described in the preceding section. Following the approaches in \cite{haus_mode-locking_2000, rana_quantum_2004, perego_coherent_2020, averchenko_effect_2024}, the dynamics of the slowly varying field amplitude $ A(t, T)$ are captured by the Heisenberg-Langevin equation for the slowly varying envelope of a pulse inside the cavity:
	\begin{align}
	\frac{\partial  A(t,T)}{\partial T} 
	&= \Bigg(-\frac{\gamma}{2} + i \Delta - i D \frac{\partial^2}{\partial t^2}\Bigg)  A(t,T)
     +  \CT  A_{\text{in}}(t,T).
        \label{eq: sec 2: SPOPO general}
	\end{align}
Here, $\gamma$ is the decay rate of the cavity.  $\Delta$ represents a detuning, defined as the round-trip phase shift of the pulse carrier frequency. For simplicity, we assume  the carrier frequency is resonant with the cavity mode, meaning the round-trip phase shift is an integer multiple of $2\pi$, or $\Delta T_R=k(\w_0)L=2\pi  n$ for $n\in N$. This resonant condition effectively sets $\Delta =0$, allowing us to disregard the detuning term in the following analysis.
The variable $t$  corresponds to the local time, which describes the pulse envelope's shape. An additional time variable $T$ represents a coarse time-scale, that tracks the pulse amplitude on a slow temporal grid, defined in multiples of the cavity round-trip time $T_R$ \cite{averchenko_quantum_2011,jiang_timefrequency_2012,jankowski_ultrafast_2024}. Specifically, $T=NT_R$ captures the state of the pulse  after $N$ round-trips.  This continuous variable $T$ is the physical counterpart to the discrete index $i$ used in the equation \eqref{eq: sec 1: spectral amplitude}.
GVD is quantified by a parameter $D = \sqrt{\cal R} k''(\w_0)L/ 2 T_R$.  The last term $ A_{\text{in}}(t,T)$ represents the input light field.  

The decomposition \cite{blow_continuum_1990, brecht_photon_2015, raymer_temporal_2020} of the amplitude $ A(t,T)$ into a series of the Hermite-Gaussian modes $\{s_n(t)\}$ yields the following expression for the $n$th mode:
\begin{align}
    \frac{\partial  a_n(T)}{\partial T}  &=  -\frac{\gamma_n}{2}  a_n(T) - i \sum_{m}C_{nm}  a_m(T) +  f_{n} (T).  \label{eq: sec 2: nth mode field}
\end{align}
Here, $\gamma_n$ is the decay rate of the $n$th mode. The coefficients $C_{nm}$ define the dispersive coupling between different modes ($n \neq m$) and the dispersive detuning for individual modes ($n = m$). Their values are calculated using the next relation:
\begin{align}
    2\frac{C_{nm}}{\gamma_n}&=\frac{N_\gamma}{N_D} O_{nm}, \label{eq: sec 2: Cnm}
\end{align}
where $O_{nm}$ is the coefficient given in equation \eqref{eq: sec 1: coefficients Onm}. $N_\gamma$ and $N_D$ are defined as previously.
%

%

%
To find a stationary solution, the time derivative in \eqref{eq: sec 2: nth mode field} is set to zero. A subsequent Fourier transform then yields the following linear system of equations:
\begin{align}
    0  &=  -\frac{\gamma_n}{2}  a_n(\Omega)  - i \sum_{m}C_{nm}  a_m(\W) +   f_{n} (\W). \label{eq: sec 2: steady state coupled equation}
\end{align}
For brevity, the dependence on $\W$ will be omitted from all subsequent equations.
The system of coupled equations can be solved using perturbation theory (PT), treating the mode coupling coefficients $C_{nm}$ as the perturbation parameters. In the zeroth-order approximation of PT, the coefficients $C_{nm}$ are set to zero, decoupling the system. The resulting steady-state solution is:
 \begin{align}
      a_n^{(0)} = 2 f_n^{(0)}/\gamma_n.
     \label{eq: sec 2: the 0th order solution} 
 \end{align}
Therefore, in the zeroth-order of PT, the amplitude of the $n$th intracavity mode is determined solely by its corresponding external field.
Substituting the zeroth-order solution  \eqref{eq: sec 2: the 0th order solution} into the linear coupling term of the steady-state equations  \eqref{eq: sec 2: steady state coupled equation} yields the first-order perturbation equations. These first-order equations are structurally identical to their zeroth-order counterparts, differing only in the form of their noise term.
\begin{align}
    0 &= -\frac{\gamma_n}{2}  a^{(1)}_n(\Omega) +  f^{(1)}_{n},\label{eq: sec 2: 1st order equation}\\ 
   &\text{where}\;\;\;  f_n^{(1)}=  f_n^{(0)} - i\sum_{m}C_{nm}  a_m^{(0)}.\nonumber
\end{align}
Hence, by combining \eqref{eq: sec 2: the 0th order solution} with \eqref{eq: sec 2: 1st order equation},  the amplitude of the $n$th in first-order PT mode is given by
 \begin{align}
     a_n^{(1)} = \frac{2 f_n^{(0)}}{\gamma_n} - i \sum_{m}C_{nm} \frac{4 f_m^{(0)}}{\gamma_n^2}. \label{eq: sec 2: the 1st order solution}
 \end{align}
The second-order PT equations are obtained by iterating the previous step. This involves substituting the first-order solution from \eqref{eq: sec 2: the 1st order solution} into the steady-state equations \eqref{eq: sec 2: steady state coupled equation}:
\begin{align}
    0 &= -\frac{\gamma_n}{2}  a^{(2)}_n(\Omega) +   f^{(2)}_{n},\\
     &\text{where}\;\;\;  f_n^{(2)}= f_n^{(0)} - i\sum_{m}C_{nm}  a_m^{(1)}.
\end{align}
Solving these equations yields the second-order PT expression for the amplitude of the $n$th mode.
\begin{align}
      a_n^{(2)} = \frac{2 f_n^{(0)}}{\gamma_n} - i \sum_{m}\frac{4 C_{nm}}{\gamma_n^2}  f_m^{(0)}  - \sum_{m,p}\frac{8 C_{nm} C_{mp}}{\gamma_n^3}  f_p^{(0)}. \label{eq: sec 2: the 2nd order solution dimensional}
 \end{align}
The solution for higher orders can be found systematically by repeating this process. However, a more elegant, general solution emerges by drawing a formal analogy with the classical approach of Maclaurin series expansion. To illustrate this, let us consider the case where only the $k$th mode is pumped. Under this condition, the mode's steady-state intracavity amplitude undergoes a dispersive transformation, described by:
\begin{align}
     a_k^{\text{cav}} &= \Bigg(1 -i \frac{N_\gamma}{N_D}  O_{kk}- \Bigg(\frac{N_\gamma}{N_D} \Bigg)^2 \sum_{m}  O_{km}  O_{mk}\Bigg) a_k^{in}. 
\label{eq: sec 2: quantum solution k-k}
\end{align}
Here, we substitute relation \eqref{eq: sec 2: Cnm} into \eqref{eq: sec 2: the 2nd order solution dimensional} and redefine the pump pulse as $2 f_k^{(0)}/\gamma_k= a_k^{in}$.
Simultaneously, a portion of the photons from the $k$th mode is transferred to other modes via dispersive coupling. This process is described by the following expression for the amplitudes $ a_l$ (where $l \neq k$):
\begin{align}
     a_l^{\text{cav}} &= \Bigg( - i\Bigg(\frac{N_\gamma}{N_D} \Bigg)  O_{lk}  -\Bigg(\frac{N_\gamma}{N_D} \Bigg)^2 \sum_{m}  O_{lm} O_{mk}\Bigg) a_k^{in}.
\label{eq: sec 2: quantum solution l-k}
\end{align}
Equations \eqref{eq: sec 2: quantum solution k-k} and \eqref{eq: sec 2: quantum solution l-k} are structurally identical to \eqref{eq: sec 1: amp_k cav - amp_k in} and \eqref{eq: sec 1: amp_l cav - amp_k in}, up to the second order. Consequently, the general solution for the intracavity light field amplitudes at any order of PT retains the same matrix structure as the solution given in \eqref{eq: sec 1: the matrix solution}, i.e.
\begin{align}
    \boldsymbol{ a}^{\text{cav}} = \sum_{M=0}^{\infty}\Bigg( - i\frac{\hat N_{\gamma}} {N_D}  \hat O\Bigg)^M \boldsymbol{ a}^{\text{in}}
\label{eq: sec 2: the matrix quantum solution}
\end{align}
where  $\boldsymbol{ a}^{\text{in}}$ is the vector of input mode amplitudes entering the optical cavity, and $\boldsymbol{ a}^{\text{cav}}$ is the vector of the mode amplitudes circulating inside it.
\section{Assessing the Limits of Perturbation Theory }
\label{Sec: Comparison of the two approaches and Limits of PT}
\begin{figure}[htbp]
    \centering
    \includegraphics[width=\columnwidth]{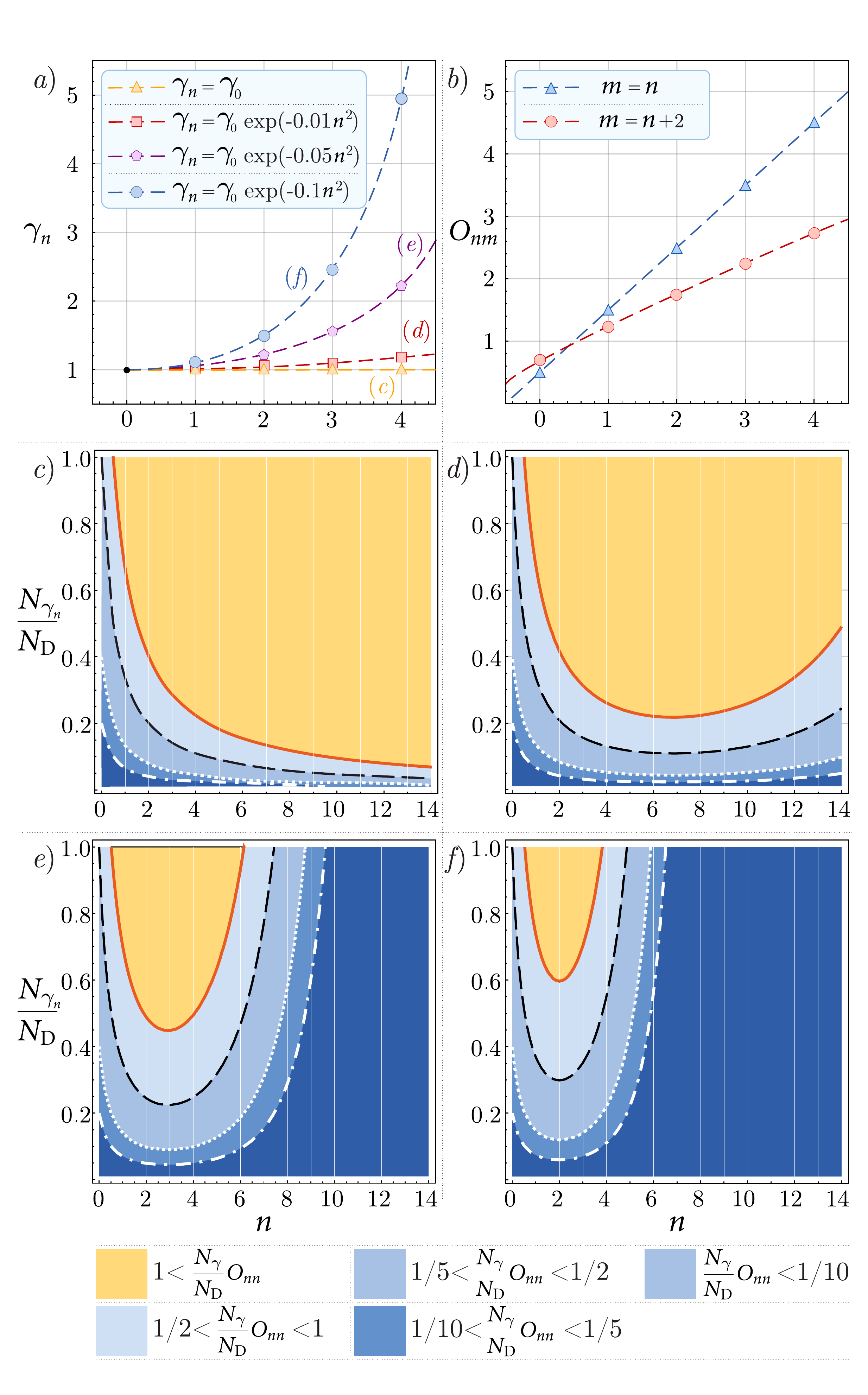}
    \caption{a) The examined dependencies of $\gamma_n$; b) The non-zero coefficient of the coupling matrix O.
        c-f), the orange-shaded regions indicate where the inequality is violated. The solid line marks the boundary where the $(N_\gamma/N_D)O_{nn}$ ratio equals 1. The dashed, dotted, and solid curves correspond to boundaries where the first term of the perturbation theory is twice, five times, and ten times larger than the second term, respectively.
    }
        \label{Fig: parameters}
    \end{figure}

The solution \eqref{eq: sec 1: the matrix solution} for light field amplitudes, based on the Maclaurin series, and the solution  \eqref{eq: sec 2: the matrix quantum solution}, derived via PT, share an identical structure. This structural equivalence implies that the system parameter condition for each mode  \eqref{eq: sec 1: main condition} must be valid in both approaches. Consequently, this correspondence provides a basis for assessing the limitations of the PT approach. 
In particular, the derivation of solutions via PT based on the assumption that the coupling coefficients $\{C_{nm}\}$ are sufficiently small. This assumption, however, is challenged by the  expression \eqref{eq: sec 1: coefficients Onm}, which reveal that the related coefficients $\{O_{nm}\}$ increase with the mode order $n$. Specifically, the diagonal elements $\{O_{nn}\}$ -- one of the central components in condition \eqref{eq:  sec 1: main condition} -- scales linearly with $n$ (see Fig.\ref{Fig: parameters}b). Consequently, if the ratio $\hat N_\gamma/N_D$ remains constant with $n$, the validity of perturbation theory is necessarily limited.  It breaks down for all modes whose order exceeds a critical threshold $n_{lim} = \lceil (2 N_D - N_\gamma) /2 N_\gamma \rceil$.
However, the potential impact of the growing coefficients $O_{nn}$ is ultimately limited by the ratio $\hat N_\gamma/N_D$, ensuring their product can satisfy the inequality. Crucially, $\hat N_{\gamma}$  is not a constant; its elements are inversely proportional to the decay rate $\gamma_n$. Therefore it is unique for each Hermite-Gaussian mode. In practice, decay rate may depend on the mode number, for example, due to finite spectral transmission of cavity mirrors. 

%

%
In our numerical analysis, we examine several dependencies of $\gamma_n$, as shown in Fig. \ref{Fig: parameters}a
. By combining these with the dependence of $O_{nn}$ on\; Fig.\ref{Fig: parameters}b, we construct contour plots Fig. \ref{Fig: parameters}(c-f) to estimate the parameters collectively and determine the value of $( N_{\gamma}/N_D)O_{nn}$.
In each figure \ref{Fig: parameters}(c-f), the orange-shaded regions indicate where the inequality is violated. The solid line marks the boundary where the $(N_\gamma/N_D)O_{nn}$ ratio equals 1. We also plot contours representing specific thresholds for this ratio. The dashed, dotted, and solid curves correspond to boundaries where the first term of the perturbation theory is twice, five times, and ten times larger than the second term, respectively.

\section{Conclusion}

This work aimed to determine the validity range of perturbative solutions for the dynamics of optical pulses propagating within a dispersive cavity. Using an empty cavity synchronously pumped by a pulsed laser as a test problem, we compared perturbative solutions to a rigorous steady-state solution. This comparison revealed a unified structure for both approaches, enabling a direct assessment of the limitations of perturbation theory.

The analysis showed that the perturbative solution remains valid only within specific parameter regimes, which depend on the interplay of dispersion effects, mode order, and cavity decay rates. Notably, higher-order modes are more susceptible to the breakdown of perturbation theory due to their stronger coupling to dispersive effects. However, the decay rate of each mode can mitigate these impacts, with modes exhibiting higher decay rates being less influenced by dispersion.
%

\section*{ACKNOWLEDGMENTS}
K.T. and D.M. acknowledge the financial support from the Russian
Science Foundation (Project No. 24-22-00318).

\appendix
\section{The solution of Sellmeier equation } \label{App: Sellemeir equation}
The Sellmeier equation for nonlinear medium  describes its refractive index $n(\lambda)$ as a function of wavelength  $\lambda$. These equations are essential for calculating dispersion parameters like GVD ($k''(\w_0)$) and TOD ($k'''(\w_0)$).

For BiBO, the Sellmeier equations are typically given in the form:
\begin{align}
n^2(\lambda) = 3.07403+\frac{0.03231}{\lambda^2-0.03163}-0.01338\lambda^2,
\end{align}
where $\lambda$ is the wavelength in micrometers ($\mu$m).

The GVD and TOD coefficients are obtained by differentiating the Sellmeier equation. The resulting expressions in terms of $\lambda$ are:

1. Group Velocity Dispersion (GVD):
  \begin{align}
   k''(\lambda) = \frac{\lambda^3}{2 \pi c^2} \frac{d^2 n}{d \lambda^2} \quad \text{(in fs$^2$/mm)}
   \end{align}

2. Third-Order Dispersion (TOD):
  \begin{align}
   k'''(\lambda) = -\frac{\lambda^4}{4 \pi^2 c^3} \left( 3 \frac{d^2 n}{d \lambda^2} + \lambda \frac{d^3 n}{d \lambda^3} \right) \quad \text{(in fs$^3$/mm)}
   \end{align}
Consequently, the calculated third-order dispersion (TOD) at the designated operating wavelength of 795 nm is:
\begin{align}
     k'''(\lambda_0)= 1644\;\; \text{fs}^3\text{mm}^{-1} .
\end{align}

\bibliography{bib}

@article{brecht_demonstration_2014,
	title = {Demonstration of coherent time-frequency Schmidt mode selection using dispersion-engineered frequency conversion},
	volume = {90},
	rights = {http://link.aps.org/licenses/aps-default-license},
	issn = {1050-2947, 1094-1622},
	url = {https://link.aps.org/doi/10.1103/PhysRevA.90.030302},
	doi = {10.1103/PhysRevA.90.030302},
	pages = {030302},
	number = {3},
	journaltitle = {Physical Review A},
	journal = {Phys. Rev. A},
	author = {Brecht, Benjamin and Eckstein, Andreas and Ricken, Raimund and Quiring, Viktor and Suche, Hubertus and Sansoni, Linda and Silberhorn, Christine},
	urldate = {2025-07-01},
	date = {2014-09-12},
	langid = {english},
}

@article{ashby_temporal_2020,
	title = {Temporal mode transformations by sequential time and frequency phase modulation for applications in quantum information science},
	volume = {28},
	issn = {1094-4087},
	url = {https://opg.optica.org/abstract.cfm?URI=oe-28-25-38376},
	doi = {10.1364/OE.410371},
	pages = {38376},
	number = {25},
	journaltitle = {Optics Express},
	journal = {Opt. Express},
	author = {Ashby, James and Thiel, Valérian and Allgaier, Markus and d’Ornellas, Peru and Davis, Alex O. C. and Smith, Brian J.},
	urldate = {2025-07-01},
	date = {2020-12-07},
	langid = {english},
}

@article{horoshko_interferometric_2024,
	title = {Interferometric sorting of temporal Hermite-Gauss modes via temporal Gouy phase},
	volume = {110},
	issn = {2469-9926, 2469-9934},
	url = {https://link.aps.org/doi/10.1103/PhysRevA.110.033721},
	doi = {10.1103/PhysRevA.110.033721},
	pages = {033721},
	number = {3},
	journaltitle = {Physical Review A},
	journal = {Phys. Rev. A},
	author = {Horoshko, Dmitri B. and Kolobov, Mikhail I.},
	urldate = {2025-07-01},
	date = {2024-09-30},
	langid = {english},
}

@article{cygan_cavity_2021,
	title = {Cavity buildup dispersion spectroscopy},
	volume = {4},
	url = {https://www.nature.com/articles/s42005-021-00517-3},
	doi = {10.1038/s42005-021-00517-3},
	pages = {14},
	number = {1},
	journal = {Commun. Phys.},
	author = {Cygan, A. and et. al.},
	date = {2021-01-29},
	langid = {english},
}

@article{averchenko_effect_2024,
	title = {Effect of group-velocity dispersion on the generation of multimode pulsed squeezed light in a synchronously pumped optical parametric oscillator},
	volume = {26},
	issn = {1367-2630},
	url = {https://iopscience.iop.org/article/10.1088/1367-2630/ad9be1},
	doi = {10.1088/1367-2630/ad9be1},
	pages = {123017},
	number = {12},
	journaltitle = {New Journal of Physics},
	journal = {New J. Phys.},
	author = {Averchenko, V A and Malyshev, D M and Tikhonov, K S},
	urldate = {2025-03-09},
	date = {2024-12-01},
}

@article{Roslund2014,
	title = {Wavelength-multiplexed quantum networks with ultrafast frequency combs},
	volume = {8},
	pages = {109--112},
	number = {2},
	journal = {Nat. Photonics},
	author = {Roslund, J. and De Araújo, R. M. and Jiang, S. and Fabre, C. and Treps, N.},
	urldate = {2024-05-23},
    doi={10.1038/nphoton.2013.340},
	year = {2014},
	}

@article{rana_quantum_2004,
	title = {Quantum Noise of Actively Mode-Locked Lasers With Dispersion and Amplitude/Phase Modulation},
	volume = {40},
	rights = {https://ieeexplore.ieee.org/Xplorehelp/downloads/license-information/{IEEE}.html},
	pages = {41--56},
	number = {1},
	journal = {{IEEE} J. Quantum Electron.},
	author = {Rana, F. and Ram, R.J. and Haus, H.A.},
	urldate = {2024-06-30},
    doi = {10.1109/JQE.2003.820831},
	year = {2004},
	langid = {english},
}

@article{jiang_timefrequency_2012,
	title = {A time/frequency quantum analysis of the light generated by synchronously pumped optical parametric oscillators},
	volume = {14},
	pages = {043006},
	number = {4},
	journal = {New J. Phys.},
	author = {Jiang, S. and Treps, N. and Fabre, C.},
	urldate = {2024-06-30},
    doi ={10.1088/1367-2630/14/4/043006},
	year = {2012},
}

@article{averchenko_quantum_2011,
	title = {Quantum correlations and fluctuations in the pulsed light produced by a synchronously pumped optical parametric oscillator below its oscillation threshold},
	volume = {61},
	pages = {207--214},
	number = {1},
	journal = {Eur. Phys. J. D},
	author = {Averchenko, V. A. and Golubev, Yu. M. and Fabre, C. and Treps, N.},
	urldate = {2024-06-30},
	year = {2011},
    doi ={10.1140/epjd/e2010-00280-7},
	langid = {english},
}

@article{brecht_photon_2015,
	title = {Photon Temporal Modes: A Complete Framework for Quantum Information Science},
	volume = {5},
	shorttitle = {Photon Temporal Modes},
	pages = {041017},
	number = {4},
	journal = {Phys. Rev. X},
	author = {Brecht, B. and Reddy, Dileep V. and Silberhorn, C. and Raymer, M. G.},
	urldate = {2024-07-02},
	year = {2015},
    doi = {10.1103/PhysRevX.5.041017},
	langid = {english},
}

@article{fabre_modes_2020,
	title = {Modes and states in quantum optics},
	volume = {92},
	number = {3},
	urldate = {2024-07-24},
	journal = {Rev.  Mod. Phys.},
	author = {Fabre, C. and Treps, N.},
	month = sep,
    doi={10.1103/RevModPhys.92.035005},
	year = {2020},
	pages = {035005},
}

@article{blow_continuum_1990,
	title = {Continuum fields in quantum optics},
	volume = {42},
	rights = {http://link.aps.org/licenses/aps-default-license},
	pages = {4102--4114},
	number = {7},
	journal = {Phys. Rev. A},
	author = {Blow, K. J. and Loudon, Rodney and Phoenix, Simon J. D. and Shepherd, T. J.},
	urldate = {2024-07-02},
    doi = {10.1103/PhysRevA.42.4102},
	year = {1990},
	langid = {english},
}

@article{lemieux_engineering_2016,
	title = {Engineering the Frequency Spectrum of Bright Squeezed Vacuum via Group Velocity Dispersion in an {SU}(1,1) Interferometer},
	volume = {117},
	pages = {183601},
	number = {18},
	journal = {Phys. Rev. Lett.},
	author = {Lemieux, Samuel and Manceau, Mathieu and Sharapova, Polina R. and Tikhonova, Olga V. and Boyd, Robert W. and Leuchs, Gerd and Chekhova, Maria V.},
	urldate = {2024-07-02},
	date = {2016-10-27},
    doi = {10.1103/PhysRevLett.117.183601},
	langid = {english},
}

@article{jankowski_ultrafast_2024,
	title = {Ultrafast second-order nonlinear photonics—from classical physics to non-Gaussian quantum dynamics: a tutorial},
	volume = {16},
	shorttitle = {Ultrafast second-order nonlinear photonics—from classical physics to non-Gaussian quantum dynamics},
	pages = {347},
	number = {2},
    doi ={10.1364/AOP.495768},
	journal = {Adv. Opt. Photonics},
	author = {Jankowski, M. and Yanagimoto, R. and Ng, E. and Hamerly, R. and {McKenna}, T. P. and Mabuchi, H. and Fejer, M. M.},
	urldate = {2024-07-05},
	year = {2024},
	langid = {english},
}

@article{gouzien_morphing_2020,
	title = {Morphing Supermodes: A Full Characterization for Enabling Multimode Quantum Optics},
	volume = {125},
	shorttitle = {Morphing Supermodes},
	pages = {103601},
	number = {10},
	journal = {Phys. Rev. Lett.},
	author = {Gouzien, E. and Tanzilli, S. and D'Auria, V. and Patera, G.},
	urldate = {2024-07-06},
    doi={10.1103/PhysRevLett.125.103601},
	year = {2020},
	langid = {english},
}

@article{perego_coherent_2020,
	title = {Coherent master equation for laser modelocking},
	volume = {11},
	pages = {311},
	number = {1},
	journal = {Nat. Commun.},
	author = {Perego, A. M. and Garbin, B. and Gustave, F. and Barland, S. and Prati, F. and De Valcárcel, G. J.},
	urldate = {2024-07-17},
	year = {2020},
    doi ={10.1038/s41467-019-14013-4},
	langid = {english},
}

@article{haus_mode-locking_2000,
	title = {Mode-locking of lasers},
	volume = {6},
	rights = {https://ieeexplore.ieee.org/Xplorehelp/downloads/license-information/{IEEE}.html},
	pages = {1173--1185},
	number = {6},
	journal = {{IEEE} J. Sel. Top. Quantum Electron.},
	author = {Haus, H.A.},
	urldate = {2024-07-17},
    doi ={10.1109/2944.902165},
	year = {2000},
}

@article{raymer_temporal_2020,
	title = {Temporal modes in quantum optics: then and now},
	volume = {95},
	shorttitle = {Temporal modes in quantum optics},
	number = {6},
	urldate = {2024-07-26},
	journal = {Physica Scripta},
	author = {Raymer, Michael G and Walmsley, Ian A},
	month = jun,
	year = {2020},
    doi ={10.1088/1402-4896/ab6153},
	pages = {064002},
}
    
\end{document}